\documentclass[aps,pre,floats,floatfix,preprint,superscriptaddress]{revtex4}

\usepackage{latexsym,amsmath}
\usepackage{amsbsy}
\usepackage[pdftex]{graphicx}
\usepackage{rotating}
\usepackage[usenames]{color}
\usepackage{amssymb}
\usepackage{multirow}
\usepackage{float}

\begin{document}

\title{Predicting effects of structural stress in a genome-reduced model bacterial metabolism}

\author{Oriol G\"uell} 
\thanks{To whom correspondence should be addressed. E-mail: oguell@ub.edu}
\affiliation{Departament de Qu\'{\i}mica F\'{\i}sica, Universitat de Barcelona, Mart\'{\i} i Franqu\`es 1,Barcelona 08028, Spain}
\author{Francesc Sagu\'es} 
\affiliation{Departament de Qu\'{\i}mica F\'{\i}sica, Universitat de Barcelona, Mart\'{\i} i Franqu\`es 1,Barcelona 08028, Spain}
\author{ M. \'Angeles Serrano}
\affiliation{Departament de F\'{\i}sica Fonamental, Universitat de Barcelona, Mart\'{\i} i Franqu\`es 1,Barcelona 08028, Spain}

\begin{abstract}
We studied in silico effects of structural stress in \emph{Mycoplasma pneumoniae}, a genome-reduced model bacterial organism, by tracking the damage propagating on its metabolic network after a deleterious perturbation. First, we analyzed failure cascades spreading from individual reactions and pairs of reactions and compared the results to those in \emph{Staphylococcus aureus} and \emph{Escherichia coli}. To alert to the potential damage caused by the failure of individual reactions, we propose a generic predictor based on local information that identifies target reactions for structural vulnerability. With respect to the simultaneous failure of pairs of reactions, we detected strong non-linear amplification effects that can be predicted by the presence of specific motifs in the intersection of single cascades. We further connected the metabolic and gene co-expression networks of \emph{M. pneumoniae} through enzyme activities, and studied the consequences of knocking out individual genes and clusters of genes. Damage caused by single gene knockouts reveals a strong correlation between genome-scale cascades of large impact and gene essentiality. At the same time, we found that genes controlling high-damage reactions tend to operate in functional isolation, as a metabolic protection mechanism. We conclude that the architecture of \emph{M. pneumoniae}, both at the level of metabolism and genome, seems to have evolved towards increased structural robustness, similarly to other more complex model bacterial organisms, despite its reduced genome size and its greater metabolic network linearity. Our approach, although motivated biochemically, is generic enough to be of potential use toward analyzing and predicting spreading of structural stress in any bipartite complex network.

\end{abstract}

\maketitle

\section{Introduction}
The architecture of complex networks is imprinted with universal features that affect their resilience and condition their behavior \cite{Albert:2002,Dorogovtsev:2008,Barrat:2008}. Most relevant, the scale-free connectivity of many natural and man-made networks explains their fragility in front of attacks to the most connected nodes, while they are able to deal with accidental failures of single components \cite{Havlin:2000,Barabasi:2000b}. Yet, current network studies mainly focus on the structural consequences of single node failures, and systemic responses to more globalized forms of structural and functional stress still remain to be explored.

In the biological context, networks of molecular interactions in the cell are among the best probed in terms of robustness in front of a variety of in silico perturbation experiments. They have been found to comply with the design principles of error-tolerant scale-free networks \cite{Barabasi:2004}, and recent progress in network dynamics is also starting to portray the concept of stress-induced network rearrangements \cite{Szalay:2007,Motter:2008}. Interestingly, metabolic networks offer an excellent arena for network stress testing and prediction, due to the amount and quality of the experimental data underlying their genome-scale reconstructions \cite{Palsson:2006} which enable reliable complex network representation and analysis \cite{Jeong:2000,Ma:2003a,Guimera:2005b,Serrano:2011b,Serrano:2012a}. In this context, the exploration of single biochemical reaction failures has shown that the structural organization of metabolic networks reduce the likelihood of large damaging cascades \cite{Smart:2008}. At the same time, many individual mutations that eliminate enzyme-coding genes seem to have very little effect on cell growth \cite{Palsson:2000,Segre:2002}. By contrast, the impact of multiple failures could go beyond the mere accumulation of individual effects, producing amplified damage due to peculiar biochemical interweaving or gene epistatic interactions \cite{Folger:2011}.

In this work, we explore and predict the effects of different forms of structural stress on the robustness of the metabolic network of {\it Mycoplasma pneumoniae}, a human pathogen that has recently been proposed as a genome-reduced model organism for bacterial and archaeal systems biology \cite{Yus:2009,Kuhner:2009,Guell:2009}. Our analysis considers the failure of single and pairs of biochemical reactions and the knockout of individual genes and clusters of genes. Although it has been suggested that the greater linearity of {\it M. pneumoniae} metabolic network and its limited redundancy might threaten its robustness \cite{Yus:2009}, we found that this organism exhibits network responses and a structural robustness that is similar to those of larger model organisms, like {\it Escherichia coli} or {\it Staphylococcus aureus}, despite its reduced genome size. For all three organisms, we show that the impact of failure cascades spreading through the metabolic network can be predicted in terms of local network motifs. In this way, targets prone to introduce structural vulnerability can be readily detected prior to experimental testing without expensive computations, even for larger and more complex organisms. For {\it M. pneumoniae}, we also explored the effects of single and multiple gene knockouts by coupling, through enzyme activity, its metabolic network to the experimentally extracted gene co-expression network. We observed that genes related to high-damage reactions are essential for the organism and that their expression tends to be isolated from that of other genes. This hints at the interplay between metabolism and genome, apparently evolved to favor the robustness of this organism by avoiding the potentially catastrophic effect of coupling the co-expression of structurally vulnerable metabolic genes.

\section{Predicting damage spreading in metabolic structure}
We modeled metabolic networks as bipartite graphs with two different types of nodes, metabolites and reactions, connected by directed links. For irreversible reactions, the directionality of the flux determines the directionality of the links, while reversible reactions are treated as two coupled reactions to account for the forward and reverse fluxes. The failure of a reaction may turn some of its metabolites inviable if they cannot be maintained anymore at steady non-zero concentration. This happens when they are left without producing or consuming reactions (except for metabolites that the organisms exchange with the environment) or, in topological terms, without incoming or outgoing connections. On its turn, an inviable metabolite makes non-operational all associated reactions. This mechanism \cite{Smart:2008} propagates a cascade of structural failure that stops when all reactions remaining in the network are viable. At this point, the corresponding damage is quantified as the number of reactions turned non-operational (see Supporting Information).

Next, we present the distribution of damages for cascades triggered by individual and by pairs of reactions in the metabolic network of {\it M. pneumoniae}, as compared to {\it S. aureus} and {\it E. coli} (in Materials and Methods we provide generic characteristics extracted from the databases for the three considered organisms). We later identify network motifs responsible for the propagation of cascades, and propose a local predictor for damage.

\subsection{Genome-scale impact of individual reactions failures}
Although close to $50\%$ of all individual reaction failures in the three organisms considered did not propagate cascades, and most cascades were indeed small ($59\%$ of the cascades in {\it M. pneumoniae}, $38\%$ in {\it S. aureus}, and $55\%$ in {\it E. coli} propagate to only one or two reactions), the individual failure of some particular reactions may trigger relatively far reaching damages. This is shown in Figs. 1a-c, that display the cumulative probability distributions $P(d'_r\ge d_r)$ that the failure of a reaction $r$ attains at least $d_r-1$ other reactions in each metabolic network (see also Supporting Information). All species show similar broad distributions, although the crossover in the tail of the distribution from power-law-like to exponential-like is not evident in {\it M. pneumoniae} probably due to its limited size (see Materials and Methods). Nonetheless, by comparing with the randomized metabolic network counterparts taken as null models (see Supporting Information), all three organisms, including {\it M. pneumoniae}, show that their metabolic organization appear to have evolved towards reducing the likelihood of large failure cascades, probably lethal, or equivalently towards increased structural robustness \cite{Smart:2008}.

Along with topological structure, biochemical insight contributes to explain why some reactions trigger particularly large cascades. For {\it M. pneumoniae}, the most vulnerable reactions at the top of the damage ranking (see Supporting Information) can be classified into four groups related to vital functions. One group is associated to metabolites phosphoenolpyruvate and protein L-histidine, each solely produced by one generating reaction and both of them directly related to phosphorylation processes, vital for instance in the synthesis of ATP. The second group relates to formate, which has a prominent role in the energy metabolism on many bacteria. The third group involves reactions where the important metabolite is thioredoxin, an antioxidant protein essential to reduce oxidized metabolites, along $\text{NADP}^{+}$. Finally, the failure of reactions in the fourth group trigger large cascades that affect the synthesis of fatty acids by turning acyl carrier proteins inviable.

Prediction of the damage caused by the failure of individual reactions is possible on the basis of local information relative to the triggering reaction alone. We give the explicit mathematical expression of our predictor $P_r$ and a detailed explanation in Materials and Methods. In simple terms, reactions need to be linked to propagator motifs in order to ignite damaging cascades. Basically, these motifs are represented by branched metabolites with just one in or out connection that happens to be attached to the triggering reaction. The higher the branching ratio of these metabolites, the higher the likelihood that the reaction propagates a large cascade, and thus to become a target for structural vulnerability of the network. To give an example, the two most vulnerable reactions in {\it M. pneumoniae} produce phosphoenolpyruvate, a compound involved in glycolysis and gluconeogenesis that acts as a source of energy. It happens to be a highly-branched cascade propagator motif connected to two reversible reactions and, as a product, to eight irreversible reactions. See Supporting Information for a categorization of cascade propagator motifs in bipartite semidirected networks.

To check the predictive power of our predictor $P_r$, we measured the Spearman's rank correlation coefficient $\rho_S$ between predictors and damages for each organism. Basically, Spearman's correlation is the Pearson correlation coefficient between two ranks, here given by the positions in ordered lists of reactions according to predictor values $P_{r}$ and damages $d_r$. A high ranking position by predictor value is expected to correlate with vulnerable reactions, at the top of the damage ranking. For all three organisms, we found very high values of the correlation coefficient, which are statistically significant (see Fig. 1d). This evidences the ability of our predictor, calculated on the basis of local information, to rank reactions by damage without directly computing the effect of the failure.

\subsection {Non-linear effects in cascades triggered by pairs of reactions}
As expected, the simultaneous failure of two reactions leads to higher damages as shown in Figs. 2a-c. The graphs display the cumulative probability distributions $P(d'_{rr'} \ge d_{rr'})$ calculated from all possible pairs of reactions initiating the cascades. It is worth stressing that the order of initiation is irrelevant. Notice that the exponential cut-off is still present, and becomes more prominent even for {\it M. pneumoniae}.

For all organisms, we observed that the cascades caused by individual reactions combine in different ways when two reactions fail simultaneously (see the illustrative sketches in Fig.3, panels 3a-d). The crucial concept here is that of the pattern of interference of the respective areas of influence of the two individually considered cascades. By that we refer to all metabolites and reactions altered \footnote{By reactions altered but not removed, we mean reversible reactions that become directed by effect of the cascade.}, removed or not, by each single cascade. If there is no interference, the total damage $d_{rr'}$ is additive and equal to the sum of the two single damages $d_r$ and $d_{r'}$. Otherwise, different situations are possible leading to a combined damage that can be equal, larger or smaller than the single added values. The latter case is a univocal signature of cascade overlapping $o_{rr'}$, pointing to the existence of a common subset of reactions that fail in both cascades (the most extreme realization is when one cascade is totally contained in the other). More interesting is the situation when, irrespectively of the presence or absence of overlap, we detect a non-linearly amplified damage involving a number $a_{rr'}$ of new reactions that break down under simultaneous black outs. For all cascades, $d_{rr'}=d_{r}+d_{r'}-o_{rr'}+a_{rr'}$. Interference without amplification is the most common situation, followed by the absence of interference (Fig. 2d). In contrast, overlap and amplification happen for a very small fraction of all double cascades, and their occurrence decreases with the size of the organism (Fig. 2e). In particular, the reduced incidence of amplification represents a new signature that organizational principles at play ensure the robustness of the organisms, despite increasing complexity and interweaving.

However, amplification may have a very large impact when it occurs. For instance, pyruvate (a product of glucose metabolism and a key intersection in several metabolic pathways) provides energy by fermentation. This process reduces pyruvate into lactate, a reaction that does not trigger any black out cascade when it fails, so $d_r=1$. At the same time, pyruvate can also be decarboxylated to produce acetyl groups, the building blocks of a large number of molecules that are synthesized in cells. The failure of the first reaction in such pathway triggers a cascade of length $d_{r'}=3$. In contrast, the simultaneous failure of both the fermentation and the reduction of pyruvate induces a large cascade of size $d_{rr'}=36$, most likely lethal. As a biological explanation, we suggest that both processes are strongly interdependent to maintain the oxidation-reduction balance when fermentation is in action.

Collateral effects offer the clue to understand this amplification phenomenon. In parallel to rendering non-operational some reactions and their corresponding metabolites, a cascade can reduce the connectivity and increase the branching ratio of other viable metabolites in its influence area. When stricken by the propagation front of a second cascade, these metabolites are susceptible of becoming inviable, further spreading the failure wave. In this way, interference is a necessary but not a sufficient condition for amplification, and a large amplification can be possible even when there is no overlap and the interference between the individual cascades is small. To predict which pairs will trigger amplification, we look at metabolites in the interference of the influence areas of the two individual cascades. We confirmed that those metabolites that remain viable after each individual cascade but become inviable when the two effects are superposed will produce amplification, propagating the double cascade to new reactions. In Fig. 3, panels 3e-j, we provide the connectivity structure of all interference cascade propagator motifs responsible for amplification.

\section{Impact of gene knockouts in metabolic structure}
Reaction failures are usually associated to the disruption of an enzyme due to knockout, inhibition, or deleterious mutation of the corresponding gene. However, enzyme multi-functionality and gene essentiality are higher in {\it M. pneumoniae} as compared to other prokaryotic bacteria, so gene malfunctioning can potentially produce an acuter stress response at the level of metabolism. To address this issue, we coupled the metabolic network of {\it M. pneumoniae} to its gene co-expression network through the activity of enzymes and we considered knockouts of individual genes and clusters of co-expressed genes. Inherent to this analysis is the potential occurrence of individual, double, or multiple cascades simultaneously. We algorithmically handled multiple knock-outs as an obvious extension of the previously considered situation of pair cascades.

\subsection{Metabolic effects of individual mutations}
Individual metabolic gene knockouts or mutations inhibit the production of catalytic enzymes and induced black outs of reactions propagating in the metabolic network as a failure cascade. From existing data,  $71\%$ of the $140$ metabolic genes in {\it M. pneumoniae} have a one-to-one relation with reactions, and $21\%$ of the genes regulate multiple reactions. Seldom the same reaction may be individually regulated by different enzymes produced by different genes, which happens for only four non-damaging reactions. More often, several genes are necessary to regulate the activity of a single reaction through an enzymatic complex. Twelve complexes codified by $26\%$ of genes regulate the activity of $10\%$ of metabolic reactions in {\it M. pneumoniae}. The removal of any of the genes involved in a complex is expected to cause the failure of the reaction controlled by the complex, which in principle may increase its vulnerability. However, we observe that almost all complexes are associated to low damage reactions, again in line with structural robustness.

To study the metabolic effects of individual mutations we simulated the knock-out of all reactions associated to the gene under consideration. As explained, most often this corresponds to one single reaction but sometimes multiple reactions are removed simultaneously. One first observation is that metabolic genes affecting vulnerable reactions will trigger large failure cascades. More interestingly, genes with large associated damages in metabolism turn out to be essential or conditionally essential for {\it M. pneumoniae} (see Table 1), with a unique exception discussed below. We use the classification in Ref. \cite{Yus:2009}, where essentiality is defined according to the measured metabolic map and the definition of a minimal medium which allows  {\it M. pneumoniae} to grow. Essential genes are those that are required for the survival of the organism, meaning that the products of the reactions that they control are essential for life and cannot be produced by alternative pathways, while conditional means that essentiality depends on the media composition available.

In fact, we have checked that all conditionally essential genes with the potential of producing high damage in the metabolism of {\it M. pneumoniae} were found to have an essential orthologue (essentiality determined by loss-of-function experiments) in {\it Mycoplasma genitalium} \cite{Yus:2009}, a comparable genome-reduced bacterium. The only exception to essentiality in Table 1 is gene {\it mpn062}, considered as non-essential in Ref. \cite{Yus:2009}, while in our study it triggers a large failure cascade and so can be classified as a vulnerable target for metabolic structure. Its damaging potential can be explained by the fact that each of the four reactions controlled by the gene has a contribution that, although not extremely high individually, adds to the total damage and interferes to produce amplification. We propose {\it mpn062} as an essential gene for metabolic function in {\it M. pneumoniae}, a conjecture that is supported by the essentiality of its orthologue in {\it M. genitalium}.

Another interesting case is essential gene mpn429, whose knockout triggers the largest cascade in {\it M. pneumoniae}. Each of the four reactions it affects in the glycolysis pathway is not able to propagate a cascade individually. However, when they all are removed simultaneously, we observe the strongest amplification effect. The biochemical explanation is that the non-linear interaction of the cascades stops the production of phosphoenolpyruvate, which disrupts the synthesis of ATP, a circumstance particularly harming to the organism.

\subsection{Metabolic effects of knocking out gene co-expression clusters}
Groups of co-expressed genes in {\it M. pneumoniae} can be identified from gene expression data under different conditions, this revealing a complex gene regulatory machinery \cite{Guell:2009}. The functional deactivation of these clusters might be produced by the failure of common regulatory elements and important damage could be transmitted to metabolism.

From this point of view, we investigated the effects on the metabolic structure of {\it M. pneumoniae} of suppressing gene co-expression clusters. To detect these functional clusters, we used three very different strategies applied to the correlation matrix of dependencies between the expression level of pairs of genes in {\it M. pneumoniae}  (see Materials and Methods and Supporting Information). We first considered clusters as defined in \cite{Guell:2009}, where a technique of average distance hierarchical clustering is used and clusters are defined as groups of nodes which are close to each other. Alternatively, we applied a random-walk-based algorithm called Infomap \cite{Rosvall:2008}, where groups comprise nodes among which information flows quickly and easily. And finally, we used the method of Recursive Percolation, that identifies clusters as strongly interconnected groups of nodes. Notice that the correlation matrix involves all genes, including those non-metabolic. Although the latter do not affect directly metabolic function, they might act indirectly due to epistatic interactions.

The comparative analysis of the detected clusters of genes showed that, although the partitions found by each algorithm may differ in their composition and in the maximum size of the clusters (see Supporting Information), there are preserved commonalities independently of the method. One of them is that all methods are able to detect seven of the twelve complexes, since the related genes always appear classified in the same cluster. Another remark is that the three detection methods result in qualitatively similar power-law-like cluster size distributions (see Supporting Information), with most clusters having small size while some are relatively big. Interestingly, genes related to high damage spreading reactions are secluded into mono-component clusters. To be more precise, eight of the nineteen genes in Table 1 are recognized by all three methods as having an expression profile that is not correlated to other gene activity levels. This is surprising since, in principle, high-damage genes might be expected to be co-regulated with other genes, as influencing big parts of metabolism usually requires coordinated gene activity. The fact that these genes appear isolated point to them as potentially important metabolic regulator targets, since the alteration of only one gene may affect metabolism at large. In any case, this is again an indication that the structural organization of the organism has evolved to increase robustness by avoiding possible amplification effects that make the system more vulnerable to failures.

Taking averages for equally sized clusters, we found that knockouts of co-expression clusters produce a damage on metabolic structure that increases with the number of affected metabolic genes, except when most metabolic genes in a cluster codify an enzymatic complex regulating one reaction (abrupt kink in Fig. 4a or Fig. 4d). The damage produced by the failure of the cluster also increases when plotted against the number of associated reactions (right panels in Fig. 4). In order to discount structural effects, we compare these results with those measured on randomized versions of the metabolic network of the genome-reduced bacterium. As evidenced in Fig.4,  all cluster detection methods identify clusters that produce lower damages in the real metabolic network of {\it M. pneumoniae} as compared to the randomized network. This supports the idea that not only the structural organization of the metabolic network of {\it M. pneumoniae} has evolved towards robustness but also the regulatory machinery that controls the coupled-to-metabolism co-expression of genes.

Finally, since the three cluster detection methods propose different forms of aggregating metabolic genes, we checked whether cluster composition is relevant for failure propagation. As a null model, we considered randomization restricted not to the network itself but to the specific gene metabolic composition, while maintaining the total number of metabolic genes in each cluster. We observed that such a reshuffling of metabolic genes in clusters had no relevant effect on the damages measured on the metabolic network (see Supporting Information). This means that, surprisingly, the composition of the clusters is not as statistically relevant for metabolic vulnerability as the distribution of the cluster sizes itself. This feature, together with the large detected amount of mono-component clusters, point out to the existence of multiple levels of regulation, depending on experimental conditions, and, at the same time, explains why genes controlling high damage spreading reactions operate preferentially under functional isolation as a metabolism protection mechanism.

\section{Conclusions}

Taken together, our results shed light on the genome-scale impact and potential control of failure propagation under structural stress in bacterial metabolism, with potential implications in areas like metabolic engineering or disease treatment. In this context, we demonstrate that {\it M. pneumoniae}, in spite of its reduced and structurally simpler genome, exhibits a similar topology and robustness when compared to other model bacterial organisms like {\it E. coli}. As a matter of fact, at the level of both single and double propagation cascades we even find more similarities between {\it M. pneumoniae} and {\it E. coli} than when compared with {\ it S. aureus} (Fig. 2d is particularly illustrative). Moreover, reactions more prone to trigger metabolic failures have been identified as key participants in the regulation of energy and fatty acid synthesis, what confers biochemical contents to our structural, network-based, analysis.

The concept of cascade amplification has been for the first time formulated and interpreted, as a signature of the subtle nonlinearities underlying the structure of complex networks. Specific scenarios in {\it M. pneumoniae} have been discussed. In addition, we were singularly motivated to assess the predicting power of our formalism. In this sense, we have proposed both a predictor of damage propagation for single cascades and we have identified structural motifs underlying amplified failure patterns in situations of concurrent spreading.

On what respects to the analysis of single gene knockouts, our analysis reveals its potentiality in capturing most of the scenarios of experimentally determined lethality for {\it M. pneumoniae}. Moreover, when clustered and knocked together new trends of the complex genomic regulation of the metabolism emerge. First, the distribution of cluster sizes seems to matter more than the actual composition of the clusters. This is connected to the fact that the regulation of high-damage genes tends to appear isolated from the that of other genes, a kind of functional switch in metabolic network that at the same time acts as a kind of genetic firewall.

The study of complex systems under stress poses a number of formidable challenges critical to understand their behavior as well as towards proposing successful strategies for prediction and control. In this framework, the study of human pathogens may help to develop more sophisticated forms of identifying new and more efficient drug targets. Finally, we emphasize that both our methodology as well as the formulated predictor index and detected propagator motifs proposed here are general tools for testing structural robustness of bipartite networks under stress, in whatever context.

\appendix
\section{Data and multilevel complex network representation}
The metabolic network of {\it M. pneumoniae} \cite{Yus:2009} contains 187 reactions divided into four types (diffusion, transport, isomerization and transformation reactions), and 228 metabolites. All reactions are taken into account to reconstruct a bipartite directed network representation, where reversible reactions are treated as two coupled directed reactions in the forward and the reverse sense. We characterize the connectivity of a metabolite by its incoming, outgoing, and bidirectional degree ($k_{i}$, $k_{o}$, and $k_b$), which count the number of reactions having the metabolite as a product, as a reactant, and the connections to reversible reactions, respectively. The degree of a reaction is the number of reactants and products involved. The total average connectivity of metabolites is 3.5, and for reactions 4.3 (see Supporting Information for further analysis of the topological properties of this network).

On the other hand, the genome of {\it M. pneumoniae} \cite{Guell:2009} comprises 688 genes, 140 of which having a metabolic function. Except for one spontaneous reaction and 20 reactions with unknown regulation, these metabolic genes codify 142 enzymes that catalize all reactions in the metabolic network of this organism. Correlations in the expression of genes were measured from tilling arrays under 62 different environmental conditions as provided in \cite{Guell:2009}. The measured gene correlation matrix can be transformed into a network representation and coupled to the metabolic network of {\it M. pneumoniae} through the activity of enzymes to produce a multilevel network representation.

Data for the bipartite directed network reconstruction of the metabolism of {\it E. coli} \cite{Feist:2007} where downloaded from the BiGG database (http://bigg.ucsd.edu/). It contains 2077 reactions and 1669 metabolites. The same criteria as for {\it M. pneumoniae} have been used to reconstruct the bipartite directed network. The average connectivity of reactions is 4.3, whereas that of metabolites is 5.3. The bipartite directed network reconstruction of {\it S. aureus} was obtained from \cite{Smart:2008}. The number of reactions is 642, after excluding sink to cytosol reactions, and the number of metabolites is 644. The average connectivity of both metabolites and reactions is 4.8.

\section{Damage predictor}
In order to predict the damage caused by the failure of individual reaction $r$, we look at its associated metabolites $m$ and compute
\begin{eqnarray}
P_r&\hspace{-0.4cm}=\hspace{-0.4cm}&\sum_{m \in r} \left[(k_{i} +k_{b}) \delta_{k_{o}}^0 (\delta_{k_{b}}^{1}+ \delta_{k_{b}}^{0}) (\delta_{k'_{o}-k_{o}}^1 + \delta_{k'_{b}-k_{b}}^1)  \right. \\ \nonumber
&&\hspace{0.5cm}+(k_{o}+k_{b}) \delta_{k_{i}}^0 (\delta_{k_{b}}^{1}+\delta_{k_{b}}^{0})(\delta_{k'_{i}-k_{i}}^1+\delta_{k'_{b}-k_{b}}^1)\\\nonumber
&&\hspace{0.5cm}- \left. \delta_{k_{i}}^0\delta_{k_{o}}^0\delta_{k_{b}}^1\delta_{k'_{b}-k_{b}}^1 \right].
 \label{eq:1d}
\end{eqnarray}
Degrees $k_{i}$, $k_{o}$ and $k_{b}$ respectively refer to the number of incoming, outgoing and bidirectional links of metabolite $m$ (reactant or product) associated to the triggering reaction $r$ after discounting the link used to propagate the cascade, and $k'_{i}$, $k'_{o}$ and $k'_{b}$ denote the same values before the cascade is triggered. We use $\delta_a^b$ for Kronecker's delta function. Basically, we consider branched metabolites attached to the triggering reaction with just one in or out connection and count their number of out and in connections, respectively.  
Bidirectional links introduce some subtleties. Some motifs involving them may propagate damage depending on the viability of the associated reversible reaction (see Supporting Information). We always take into account these motifs when we compute the value of the predictor for each reaction.

\section{Methods for detecting gene co-expression clusters}
The matrix of correlations between the expression levels of pairs of genes gives a fully connected network where the link between two genes carries a weight equivalent to the expression correlation between them. We used three different methods to detect gene co-expression clusters. First, we used the results in \cite{Guell:2009}, where a distance hierarchical clustering technique was applied to the matrix of correlations after applying a threshold of 0.65 to it (which reduces the density of links to 0.007) and transforming the Pearson correlations of expression levels into a distance. Second, we applied an existing algorithm to find communities called Infomap (Ref. \cite{Rosvall:2008}) to the matrix of correlations after applying the threshold of 0.65 to the weights. This algorithm detects communities using a random walk with jumps. A community is a set of nodes where the random walker flows quickly and easily. Third, we consider the clusters in which the co-expression network is fragmented just below the percolation threshold, where the connected network disaggregates into smaller components (other fully connected networks have been analyzed at the percolation point, see for instance \cite{Rozenfeld:2008}). To find them, we removed links sequentially from lower to higher weight until we detected the percolation transition (by computing magnitudes which have a singularity in this point, like the size of the second largest cluster and the average size of the clusters excluding the largest one). At this point we measured the clusters using a burning algorithm. Typically, the gene co-expression network fragmented into two large clusters and several small clusters.  We applied the same procedure to the two largest components and so on until the distribution of sizes was similar to those for the hierarchical clustering technique and Infomap (see Supporting Information). We called this procedure Recursive Percolation.

\begin{acknowledgments}
We thank Prof. Luis Serrano, Eva Yus, Marc G\"uell and Ashley Smart for kindly providing the empirical data. We also thank Georg Basler for helpful comments. This work was supported by MICINN Projects No. FIS2006-03525, FIS2010-21924-C02-01 and BFU2010-21847-C02-02; Generalitat de Catalunya grant No. 2009SGR1055; and the Ram\'on y Cajal program of the Spanish Ministry of Science.
\end{acknowledgments}

\begin{table*}
\centering
\begin{tabular}{clccc}
  \hline
  Gene&&Essentiality&Damage&Reactions\\
  \hline
 {\bf mpn429}&&yes&49&4 (1,1,1,1)\\
  mpn606&&yes&32&1 (32)\\
  mpn628&&yes&32&1 (32)\\
  mpn017&&yes&25&3 (14,1,9)\\
  mpn303&&yes&18&8 (1,1,1,1,8,1,2,3)\\
  {\it mpn062}&&{\it no}&17 &4 (6,3,2,3)\\
  mpn576&&cond&16 &2 (13,2)\\
  {\bf mpn005}&&yes&13&1 (13)\\
  {\bf mpn336}&&yes&13 &3 (4,3,6)\\
  {\bf mpn354}&&yes&13 &1 (13)\\
  {\bf mpn627}&&yes&11 &1(11)\\
  mpn066&&yes&9&4 (1,1,2,5)\\
  {\bf mpn240}&&cond&9&1 (9)\\
  mpn299&&cond&9&1 (9)\\
  mpn322&\multirow{3}{*}{\hspace{-0.5cm} \begin{math} \left. \begin{array}{l} \\ \\ \\ \end{array} \right\} \end{math} }&cond&9&4(1,1,2,1)\\
  mpn323&&cond&9&4 (1,1,2,1)\\
  mpn324&&cond&9&4 (1,1,2,1)\\
  {\bf mpn034}&\multirow{2}{*}{\hspace{-0.5cm} \begin{math} \left. \begin{array}{l} \\ \\ \end{array} \right\} \end{math} }&yes&7&4 (1,1,2,3)\\
  {\bf mpn378}&&yes&7&4 (1,1,2,3)\\
  \hline
 \end{tabular}
 \caption{Largest structural damages produced in metabolism by gene knockouts and correspondence with gene essentiality as given in Ref. \cite{Guell:2009}. Damage in metabolic structure caused by gene knockout (third column) is measured in number of deleted reactions. In the fourth column, we give the number of reactions regulated by the corresponding gene, and in parentheses we give the damage associated to each of these reactions. Genes in monocomponent clusters are highlighted in boldface, and we used braces to denote genes that form complexes. Note that the complex at the end of the list is not detected by any of the three clustering procedures. Finally, gene {\it mpn062} is the only one in the table annotated as non-essential although it is associated to a large failure cascade.}
 \label{table1}
\end{table*}

\begin{figure*}
\begin{center}
\includegraphics[width=0.8\textwidth]{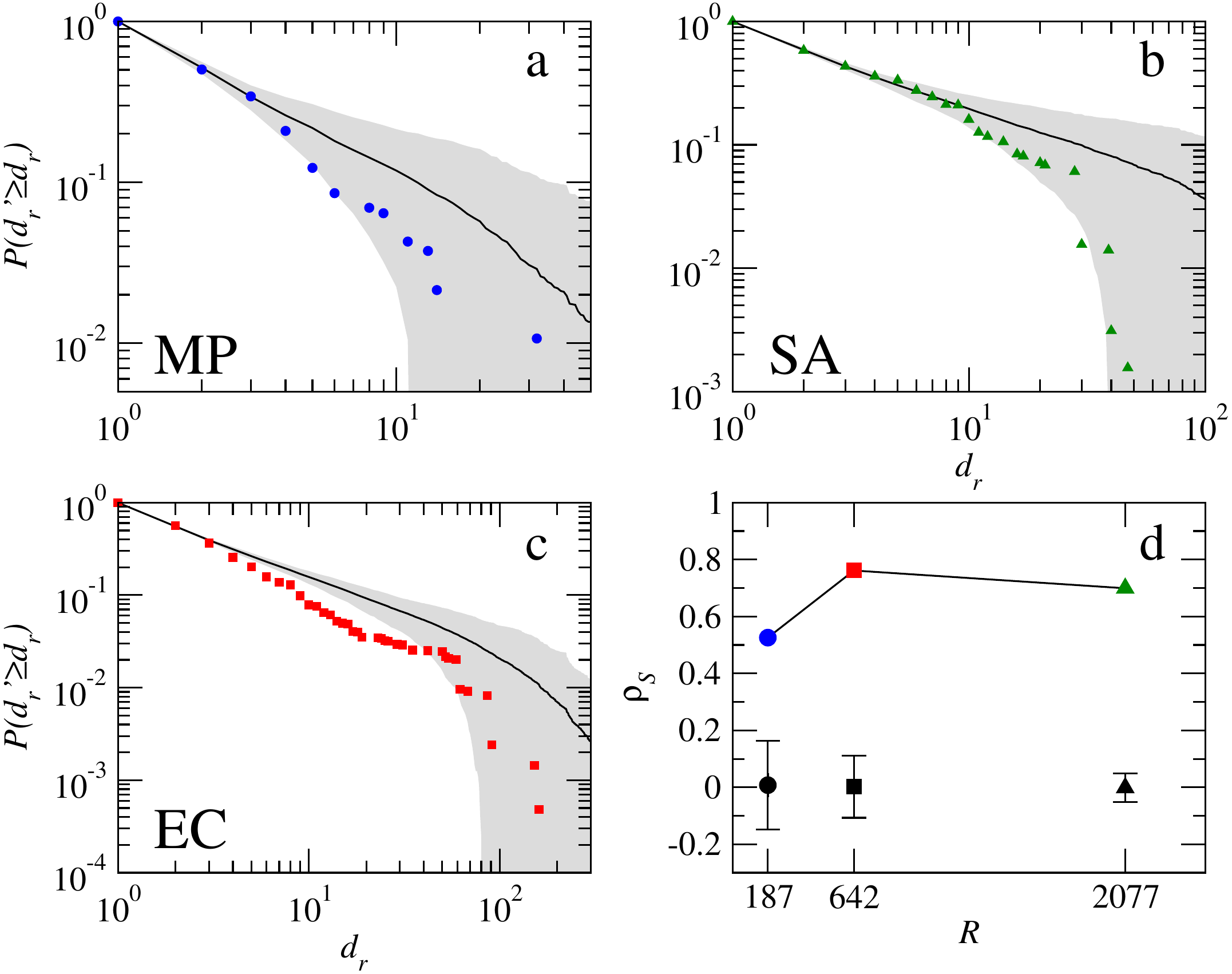}
\end{center}
\caption{Damage in cascades triggered by individual reactions. a-c). Cumulative probability distribution functions of damages in {\it M. pneumoniae}, {\it E. coli}, and {\it S. aureus}. Results are compared with damages produced in randomized versions of the metabolic networks in order to discount structural effects. In each case, the solid black curve is the average over $100$ realizations and the area in gray correspond to this average $\pm 1.96$ standard deviations. d) Spearman's rank correlation coefficient $\rho_S$ between predictors and damages, plotted against metabolic network size (number of reactions $R$). Results are compared to random reshuffling of the predictor value associated to reactions ($100$ realizations for each organism). Average Spearman's rank correlation coefficients for the randomizations appear in black, and error bars delimit the maximum and the minimum values obtained.}
\label{fig:1}
\end{figure*}

\begin{figure*}
\begin{center}
\includegraphics[width=0.8\textwidth]{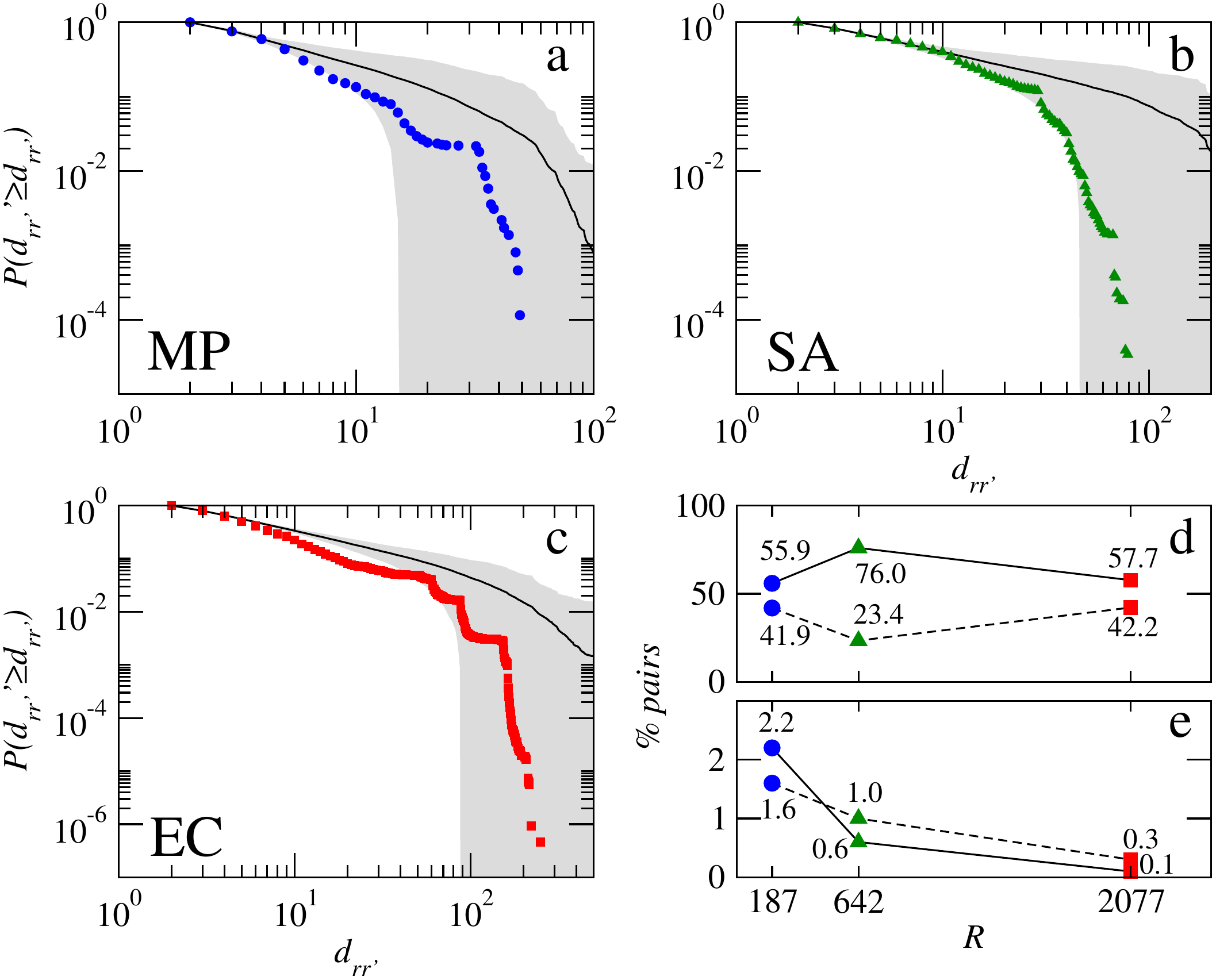}
\end{center}
\caption{Damage in cascades triggered by pairs of reactions. a-c). Cumulative probability distribution functions of damages in  {\it M. pneumoniae}, {\it E. coli}, and {\it S. aureus}. Results are compared with damages produced in randomized versions of the metabolic networks in order to discount structural effects. In each case, the solid black curve is the average over $100$ realizations and the area in gray corresponds to this average $\pm 1.96$ standard deviations. d) Most frequent double cascades output. Solid line: interference without amplification. It is related with cases b and c in Figure \ref{fig:3}. Dashed line: no interference, which is related wih case a in Figure \ref{fig:3}. e) Non-linear effects in double cascades. Solid line: overlap. It is related with cases c and e in Figure \ref{fig:3}. Dashed line: amplification. Amplification is related with cases d and e in Fig. \ref{fig:3}.}
\label{fig:2}
\end{figure*}

\begin{figure*}
\begin{center}
\includegraphics[width=0.48\textwidth]{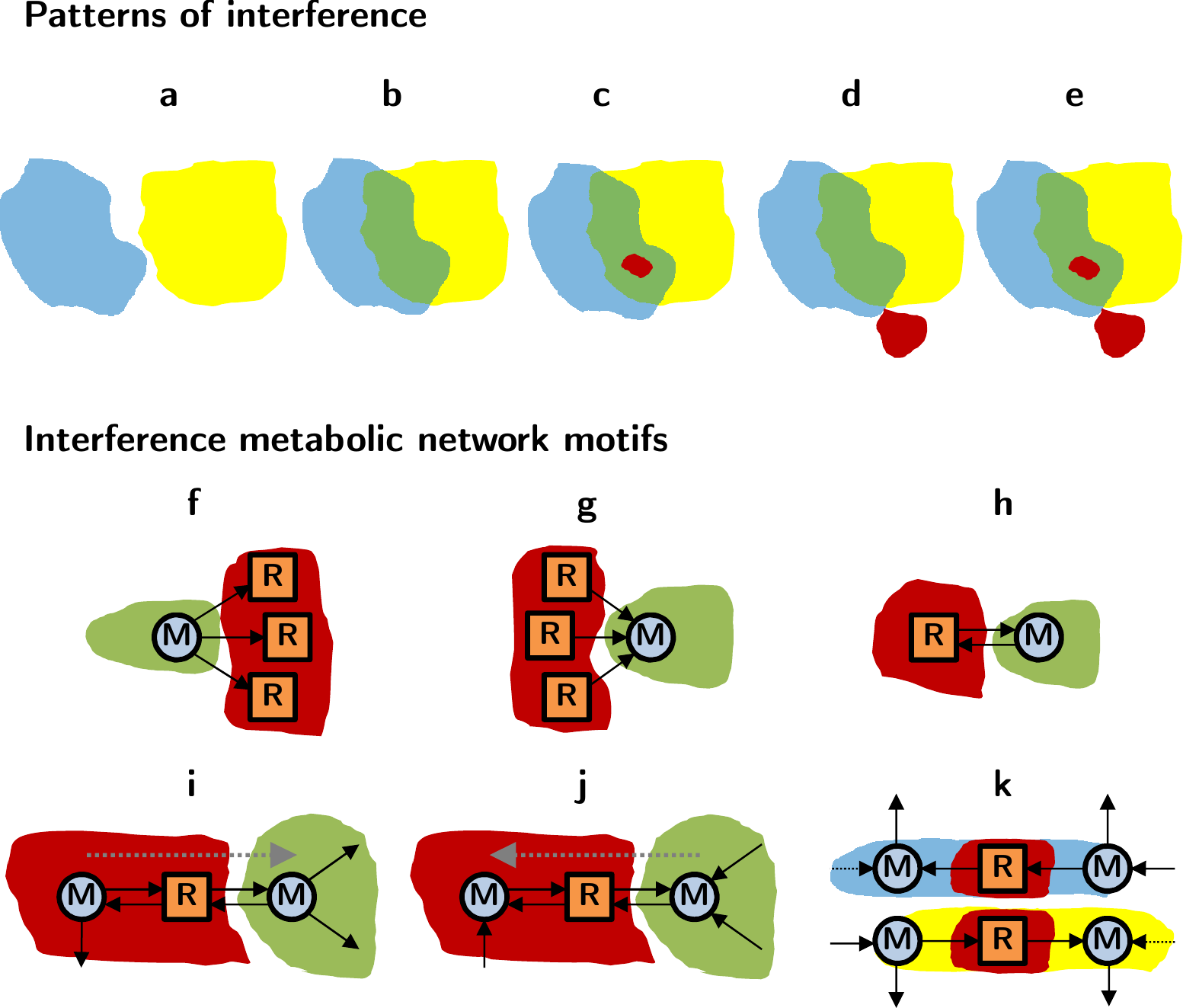}
\end{center}
\caption{Cascade propagator network motifs and typology of double cascades. a-e) Illustration of possible interference patterns between individual cascades: additive, interference without overlap or amplification, interference with overlap and without amplification, interference without overlap and with amplification, interference with overlap and amplification, respectively. Blue and yellow stand for single cascades, green for interference, and red for overlap and amplification, depending on whether the red zone is in the interference zone (green) or not. f-k) Metabolic network motifs in the interference of two individual cascades that induce amplification. Cases f-g) Motif caused by a metabolite which loses its only generating reaction and at the same time it is the reactant of several reactions. These reactions are going to be become non-viable. Case g is equivalent to f but inverting the sense of the links. Case h) Metabolite which has been left with one connection to a reversible reaction. This reversible reaction has zero net flux and becomes inviable. Cases i-j) This motif appears when a modified metabolite is lead with only one incoming connection coming from a reversible direction. This fixes the reversible reaction towards the production of this metabolite. If this step turns a metabolite of the reversible reaction inviable, the reversible reaction becomes inviable. Therefore, this motif is a potential trigger of amplification. Case j is equivalent to case i when the senses of the reactions are inverted. Case k) The individual cascades fix the sense of a reversible reaction oppositely, one cascade forwards (k top) and the other backwards (k bottom) (note that the pictures illustrate the effects of both cascades individually). After superimposing the effects of the two cascades, one can see that this reversible reaction is becomes inviable. Thus, metabolites surrounding the reaction may become inviable as well, depending on their degrees. It is also a potential trigger, as in cases i-j.}
\label{fig:3}
\end{figure*}

\begin{figure*}
\begin{center}
\includegraphics[width=0.8\textwidth]{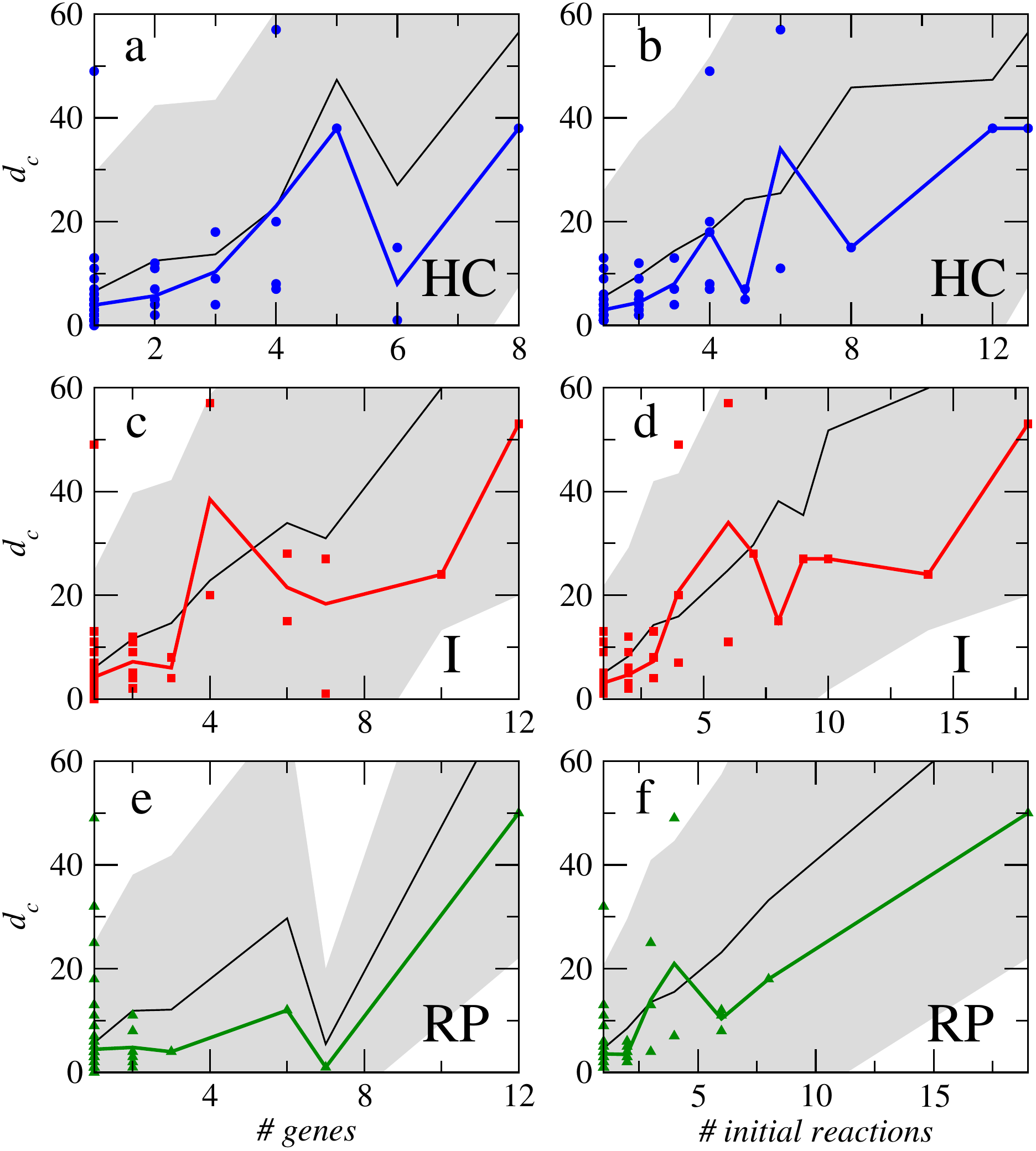}
\end{center}
\caption{Damages as a function of the number of metabolic genes and reaction failures in gene co-expression cluster knockouts. Clusters are defined according to three different methods: Hierarchical Clustering (HC), Infomap (I), and Recursive Percolation (RP), see Materials and Methods. Results are compared with damages produced in randomized versions of the metabolic networks in order to discount structural effects. In each case, the solid black curve is the average over $100$ realizations and the area in gray correspond to this average $\pm 1.96$ standard deviations.}
\label{fig:4}
\end{figure*}

\end{document}